**Title** Photoluminescence and Electron Spin Resonance of Silicon Dioxide Crystal with Rutile Structure (Stishovite)

*Anatoly Trukhin*, Andris Antuzevics*

Prof. A.N. Trukhin, Ph.D. A. Antuzevics
Institute of Solid State Physics, University of Latvia, LV-1063 Riga, Latvia
e-mail: truhins@cfi.lu.lv

Abstract text.

An electron spin resonance (ESR) and photoluminescence signal is observed in the as grown single crystal of stishovite indicating the presence of defects in the non-irradiated sample. Photoluminescence of the as received stishovite single crystals exhibits two main bands – a blue at 3 eV and an UV at 4.75 eV. Luminescence is excited in the range of optical transparency of stishovite (below 8.75 eV) and, therefore, is ascribed to defects. A wide range of decay kinetics under a pulsed excitation is observed. For the blue band besides the exponential decay with a time constant of about 18 µs an additional ms component is revealed. For the UV band besides the fast component with a time constant of 1-3 ns a component with a decay in tens µs is obtained. The main components (18 µs and 1-3 ns) possess a typical intra-center transition intensity thermal quenching.

The effect of the additional slow component is related to the presence of OH groups and/or carbon molecular defects modifying the luminescence center. The additional slow components exhibit wave-like thermal dependences. Photo-thermally stimulated creation-destruction of complex comprising host defect - interstitial modifiers explains slow luminescence wave-like thermal dependences.

1. Introduction

Silicon dioxide can exist in many polymorph modifications. Polymorph modifications based on silicon $sp^3$ hybridization comprise a family of tetrahedron structured materials, which could be considered as one of the most investigated systems from fundamental and from practical points of view. The dense (4.28 $g \cdot cm^{-3}$) octahedron structured polymorph modification named stishovite, however, is not yet well studied. The remarkable property of stishovite is its hardness (29.5 GPa) in comparison to α-quartz (9.81 GPa).[1] Also, unlike quartz, stishovite does not react with fluoric acid. On the other hand, stishovite is metastable. Heating at 825 K causes an irreversible amorphization. [2]

The main structural element of the dense silicon dioxide – stishovite – corresponds to silicon hybridization – $d^2sp^3$, resulting in an octahedral surrounding of silicon with oxygen ions. Oxygen ions are threefold coordinated.

In stishovite samples (synthetic single crystals and Meteor Crater Arizona “natural” polycrystalline powder) two main luminescence bands have been observed: a blue band at about 3 eV with a time constant of about 18 μs and a UV band at about 4.75 eV with a time constant of some ns.[3-5] Besides, luminescence with series of sharp lines in the red spectral region also has been observed. It was explained as a quasi-molecular luminescence center due to the presence of carbon impurities.[4] Also, OH groups have been found in both samples in different concentrations. [4, 5] In Arizona stishovite its concentration was much higher. [5]

Decay kinetic curves at different temperatures have revealed some peculiarities in the shape of different temperature dependences under photo and x-ray excitations. [3] The effects were interpreted as luminescence center interaction with the surrounding interstitial defects, which modify properties of the luminescence center. It was assumed that nearest defects modulates radiative transition probabilities provoking deviations the kinetic curve from exponential decay. The x-ray excitation stimulates the creation of complex defect at temperature above 100 K leading to a decrease of luminescence intensity excited in electron-hole recombination.

[7] Photoexcitation does not provide an efficient charge release and, therefore, the luminescence intensity is less dependent on temperature than that under x-ray excitation.

In α-quartz crystal no similar to stishovite luminescence was observed in the as-received samples. However, in the case of oxygen deficient silica glass similar to stishovite luminescence with two bands – one slow in the visible (2.7 eV) and one fast in the UV (4.4 eV) spectral region has been known for a long time. [6] In the case of α-quartz crystal similar to stishovite luminescence can be created by dense electron beam irradiation at low temperatures or destructive neutron and γ- irradiation at room temperature. [7, 8]

Previous electron spin resonance (ESR) studies on irradiated samples of stishovite have reported a resonance at g ≈ 2.003, whereas in-depth ESR studies of paramagnetic defects in synthetic samples have identified $H^0$, $Al\text{-}O_2^{3-}$, $Ti^{3+}$, $W^{5+}$ $Cr^{3+}$ and $Cr^{5+}$ centers in stishovite. [9-12] With the exception of $Cr^{3+}$, these impurities are effective spin $S$ = 1/2 systems with anisotropic components of the $g$ tensor and characteristic hyperfine (HF) interaction with the respective nucleus. ESR studies of non-irradiated as received stishovite crystal were not found in the literature.

Previously we have investigated the luminescence kinetics of stishovite mainly with a photomultiplier (PM) in the current regime by recording the decay curves with an oscilloscope. The photon counting method is more sensitive and for a luminescence signal with a long duration it is preferable.

In the present article we report the observation of an additional slow decay component in stishovite luminescence and the observation of an ESR signal in as received crystal. Decay curves in the range of ms have been obtained for the blue band and in µs for the UV band. Spectral and thermal dependences of this additional luminescence and the newly discovered ESR have been investigated to elucidate their nature.

## 2. Experimental procedure

Single crystals of stishovite have been grown from hydrothermal solution in the $SiO_2$–$H_2O$ system under a pressure of 9–9.5 GPa within a temperature range of 1170–770 K.[13] The natural quartz crystal powder was taken as a raw material. So, it is possible that the impurities of natural quartz could be incorporated into stishovite synthetic crystal. However, purification during the crystallization could have occurred as well and the quantity of impurities in synthetic stishovite could be smaller than those in the initial natural quartz powder. The crystals were grown in a platinum container directly contacting a graphite heater, which is the source of carbon impurities in stishovite single crystal. The samples under investigation were small, optically transparent single-crystals with the dimensions of about 0.2-0.4·0.6·0.9 $mm^3$. The small size of the samples is a source of experimental difficulties. The samples were kept in copper holders covered with an indium layer containing a hole into which the samples were pressed, avoiding slits, which would let any stray light through. The excitation was made from one side of the holder and the detection – from the other, excluding the possibility of luminescence due to contamination on the surface of the holder. For ESR measurements a small crystal was merged into drop UHU-plus transparent glue, kept on thin glass tube.

Another studied sample was a shock-wave-created stishovite in powder form found at the Meteor Crater, Arizona.[18] We pressed the powder into a plate covered with indium. The dimensions of the sample surface were about 8×8 mm. All the samples have been studied previously. [3-5]

Photon counting head Hamamatsu H8259-02 is used. Luminescence decay curves were recorded by the Tektronics TDS 2022B oscilloscope and/or Picoscope 2208, each curve being averaged over 128 pulses or more. Time-resolved luminescence spectra were extracted from the decay curves measured at the specified wavelength. The corresponding decay curve was integrated over the measured time interval. ArF (193 nm wavelength), KrF (248 nm wavelength) and $F_2$ (157 nm wavelength) excimer lasers, model PSX-100 of Neweks (Estonia) with pulse energy of up to 5 mJ and pulse duration of 5 ns were used to excite luminescence.

Samples were illuminated by unfocused beams ensuring the predominance of single photon excitation. Luminescence emission was collected in a direction perpendicular to the excitation laser beam. The samples were carefully cleaned and mounted on a holder, no glue was used. Measurements were performed at 60 – 350 K sample temperatures. The lower temperature limit corresponds to the temperature of liquid nitrogen under the pump.

Luminescence was detected with the help of a grating monochromator (MCD-2) with slit width of about 1 mm corresponding to 1.5 nm spectral resolution for time resolved spectra and steady state spectra have been measured with CCD of Hamamatsu mini spectrometer C10082CAH. Luminescence measurement details are described in. [3-5] The measured curves are presented in the figures as recorded, therefore, they reflect the level of errors.

A modified X-band RE 13-06 spectrometer (microwave frequency ≈ 9.1 GHz with 0.001 GHz precision; magnetic field modulation 100 kHz) was used for ESR spectra measurements at 77 K. The ESR spectra angular variations of the single crystal sample were made around an arbitrary rotation axis at a uniform angle interval of 5 ° with an uncertainty of 1 °. For calibration purposes a DPPH reference sample was used.

## 3.Results

### 3.1 Photoluminescence

With the use of photon counting with the H8259-02 module and oscilloscope we have detected luminescence of stishovite with duration in the range of ms. The corresponding decay curves are presented below. The PL spectra, which were obtained by integration of the decay curves, are shown in **Figure 1**. Excitation was performed with excimer lasers KrF (248 nm), ArF (193 nm) and $F_2$ (157 nm). It is seen that the time-resolved PL spectra are similar under different excitation wavelengths and temperatures. They are also similar to the previously published data. [4]

Inserts in **Figure** 2 show the decay curves for different temperatures. There is an exponential decay at 80 K similar to the previous investigations (left insert). At 120 K a slow component appears, visible in the insert in the middle. The insert on the right shows decay of PL in the 800 nm range, which belongs to carbon impurity-related centers, as was interpreted previously. [5] In **Figure** 3 the PL decay curves for the high temperature range are shown. The decay is much slower than 18 μs and is non-exponential. The intensity decreases with heating up to 500 K.

In **Figure** 4 the intensity of slow PL obtained by the integration of decay curves at each temperature is demonstrated. The thermal dependence of the time-resolved 18 μs component intensity is presented as well. These dependences are completely different. The thermal dependence of the 18 μs component is monotonous without any peaks as seen for the slow component. Temperature dependences of x-ray excited luminescence of stishovite show some kind of anti-correlation with that of the slow component. In the x-ray excited luminescence there is drop at 100 K, which anti-correlates with the increase of the slow PL intensity. A shift in thermal position between slow component time constant and corresponding intensity is observed (**Figure** 4 open squares). It is explained as acceleration of electronic transitions in a complex comprising defect-modifier with increase of the temperature. It is presumed that this complex is created in photo – thermally stimulated process. A decrease in intensity corresponds to process of destruction of complex. Two peaks, perhaps, correspond to different modifiers. The decay kinetics curves corresponding to 130 K, 210 K and 270 K temperatures are presented in **Figure** 5. It is seen that at 130 K and 270 K the decay possesses intensive slow component and at 210 K its intensity is diminished. In the **Figure** 6 the decay curves for the slow PL excited with KrF laser are shown. The decay is fast at 80 K. Increase of temperature results in increase of the slow PL intensity in the ms time range as in the case of ArF laser. The μs component behaves as that excited with ArF laser, see **Figure** 6 left insert.

Low excitation power of $F_2$ laser provides mainly PL in μs time range. Its thermal dependence shows the usual increase of intensity with cooling without any wave-like peculiarities.

In **Figure** 7 the UV band detected by the photon counting method is presented. It is seen that there is a slow component detected in the μs time range. This component is absent at 80 K, similarly to slow component of the blue band, and only fast component in ns range of time is observed. The component with decay time in the ms range is not detected for the UV band. In **Figure** 8 afterglow curves after x-ray irradiation are presented for the blue and UV bands of stishovite. For afterglow a good correspondence in the kinetics of blue and UV bands is observed, which does not take place in **Figures** 3, 5, 6 for the blue band and in **Figure** 7 for the UV band photoluminescence. Afterglow is a pure recombinative process of electrons and holes and in recombination the behavior of both bands is practically identical. From this we could confirm the previous assumption that both bands belong to the same luminescence center. [3] The difference in slow kinetics for the blue and UV bands under laser excitation could mean that process of luminescence excitation is not a recombination of released electrons and holes and can be attributed to intra-center process of excitation.

### 3.2 **Electron spin resonance**

For the non-irradiated stishovite crystal an angle-dependent ESR signal was detected at 77 K. The ESR spectra at some selected orientations are shown in Figure 10. Due to the small sample dimensions and technical limitations of the apparatus the signal-to-noise ratio is poor. The resonance position roadmap is shown in Figure 11. The isotropic resonance at $g_{eff}$ = 2.003 originates from the sample holder.

In our case, the HF structure could not be resolved in the spectra, which hinders the attribution of the observed signal to a particular impurity ion. Also, no set of spin-Hamiltonian (SH) parameters from is able to reproduce the experimental ESR spectra. [10-12]

In a simplified approach, the observed resonances could be accounted for in a S = 1/2 model and axial SH:

$$H = \beta\left[ g_{\perp}\left(B_x S_x + B_y S_y\right) + g_{\parallel} B_z S_z \right] \quad (1)$$

where β is the Bohr magneton. From the magnetic field range of experimental resonances in Figure 11 it is possible to evaluate the limits of principle g values (g1 ≈ 2.06 and g2 ≈ 2.00), however, a correct assignment to the parallel and perpendicular components of the g tensor would require a precise knowledge of the sample's crystallographic axis orientations in respect to B.

## 4. Discussion

In the as-grown single crystals of stishovite photoluminescence and ESR signals of defects existing in the samples have been obtained. The defects have been created during the crystal growth and perhaps are of the same nature and belong to host material. ESR could be explained as trapped hole centers $O^-$ and $O^{2-}$ are typical defects in oxides, however, in crystals with rutile structure their maximal *g* value is usually around $g_{eff} \approx 2.03$. [15, 16] Additionally, sample irradiation with UV or X-rays may be necessary prior to the detection of EPR spectra. In the crystals of $KTiOAsO_4$ $M^{n+}$–$O^-$–$M^{(n-1)+}$ ($M^{n+}$ – lattice cation; $M^{(n-1)+}$ – impurity cation) paramagnetic centers have the $g_{||}$ component in the 2.06-2.07 region, however, these too are radiation induced defects, which are unstable at room temperature. [17] Similar principle *g* values have also been observed in non-irradiated single-crystal citrine quartz samples, where the centers were assigned as $O_2^{3-}$ type defects. [18] However, we cannot actually say more about the observed paramagnetic centers in stishovite, therefore, further investigations are required.

In the previous studies of stishovite luminescence we observed at temperatures near 290 K the luminescence decay of the blue band was strongly non-exponential. After cooling down to 80

K the main component of the decay became exponential with some peculiarities in the initial stage of decay as shown in Fig.9.[3, 4] As it can see from this figure, at 130 K some modulations appear in initial part of the decay curve (range of hundreds of ns). We ascribed these modulations to the interaction of luminescence center with the surrounding interstitial defects such as OH groups or even water molecules as well as with molecular defects related to carbon impurities. [3]

Now we performed more detailed studies of luminescence decay in longer time range and found an additional slow decay. So, besides the already studied luminescence of stishovite single crystal with a blue band (3,1 eV), characterized with the time constant of 18 µs, and a UV band (4.75 eV) with time constant in the ns range, we have found the slower components: that in the ms range for the blue band and in tens of µs for the UV band (see **Figure**s 3, 5, 6 and 7). It should be noted, that the detection of the faster luminescence was performed in “current” regime of PM. The slowest component was difficult to detect by this method. For the detection of the slow component a photon counting method was applied. Use of the photon counting regime of PM allows detection of the slow luminescence. However, the regime of photon counting incorrectly detects luminescence in the faster range because of multi photon pulse is counted as one photon pulse. As a result in the case of multiphoton pulses we observe a strongly distorted decay curve. Therefore, here only one photon pulses related kinetics in ms range of time are discussed.

We can assume that existence of the slow decay besides the strong faster component could be due to the participation of a center in electron-hole recombination processes. However, as it was shown in **Figure** 8, in the case of electron-hole recombination such slow decay should be similar for blue and UV bands, which was the case of x-ray excitation of luminescence. In spite of observation of the slow decay under excimer laser excitation the decay was different for the blue (**Figure**s 3, 5, 6) and for the UV (**Figure** 7) bands. It means that such behavior does not correspond to recombination process related to the trapping and recombination of charge carriers. A possible explanation of the observed effect of additional slow luminescence could be in the modification of the center at different temperatures. And that modified center possesses the blue band in ms range of time and UV in µs range. Different decay of blue and UV bands of modified center witness of intra-center excitation of such a complex. In pure recombination process kinetics of both bands are very similar, **Figure** 8. Possible modification could be related to the presence of OH groups and/or carbon related molecular defects, as it was assumed in the interpretation of previous data. [3] The slow decay exceeding the time constant of the center at low temperatures and wave-like thermal dependences of slow component intensity and time duration can be explained by photo- thermally stimulated creation-destruction of complex defect-modifier. Such creation of complex should be responsible for the decrease of x-ray luminescence intensity above 100 K anti-correlated with increase of slow luminescence, **Figure** 4. [3] The complex creation is photo thermally stimulated. This is seen on difference of slow PL intensity and time constant, **Figure** 4. It is explained as acceleration of transitions in excited complex with increase of the temperature. So, there are three processes leading to luminescence of stishovite. First are intra-center electronic transitions in a defect providing two luminescence bands: a one blue due to triplet-singlet transitions and an UV band due to singlet-singlet transition. In that stishovite luminescence is similar to luminescence of oxygen deficient silica glass and heavy irradiated α-quartz crystal.[3, 5] Second process is due to intra-center transitions modulated by nearest defects, possibly OH and/or carbon related defect and third one is recombination luminescence due to photo-stimulated detachment and back recombination of nearest defect which is object of actual investigation. [4] Two last processes are possible in oxygen deficient silica glass and heavy irradiated α-quartz crystal, however not yet studied.

## 5. Conclusions

The defects of stishovite are observed by ESR and luminescence methods. They have been created during the crystal growth and perhaps are of the same nature.

The discovered ESR signal does not correlate with known signals of impurity defects in stishovite. The evaluated $g$ values range from ≈ 2.00 to ≈ 2.06. Similar principle $g$ values have been observed for oxygen ion related paramagnetic centers.

Stishovite single crystal luminescence possesses an additional slower component in the ms range for blue band and in tens of µs for the UV band. The effect of additional slow component could be related to the presence of OH groups and/or carbon related molecular defects modifying the luminescence center and creating a complex defect. This complex is affected to photo- thermal creation-destruction. It is assumed that created complex provides slow luminescence excited in intra-center process.

Acknowledgements

This work was supported by ERANET MYND. Also, financial support provided by Scientific Research Project for Students and Young Researchers Nr. SJZ/2017/2 realized at the Institute of Solid State Physics, University of Latvia is greatly acknowledged. The authors express our gratitude to R.I. Mashkovtsev for help in ESR signal interpretation. The authors are appreciative to T.I. Dyuzheva, L.M. Lityagina, N.A. Bendeliani for stishovite single crystals and to K. Hubner and H.-J. Fitting for stishovite powder of Barringer Meteor Crater.

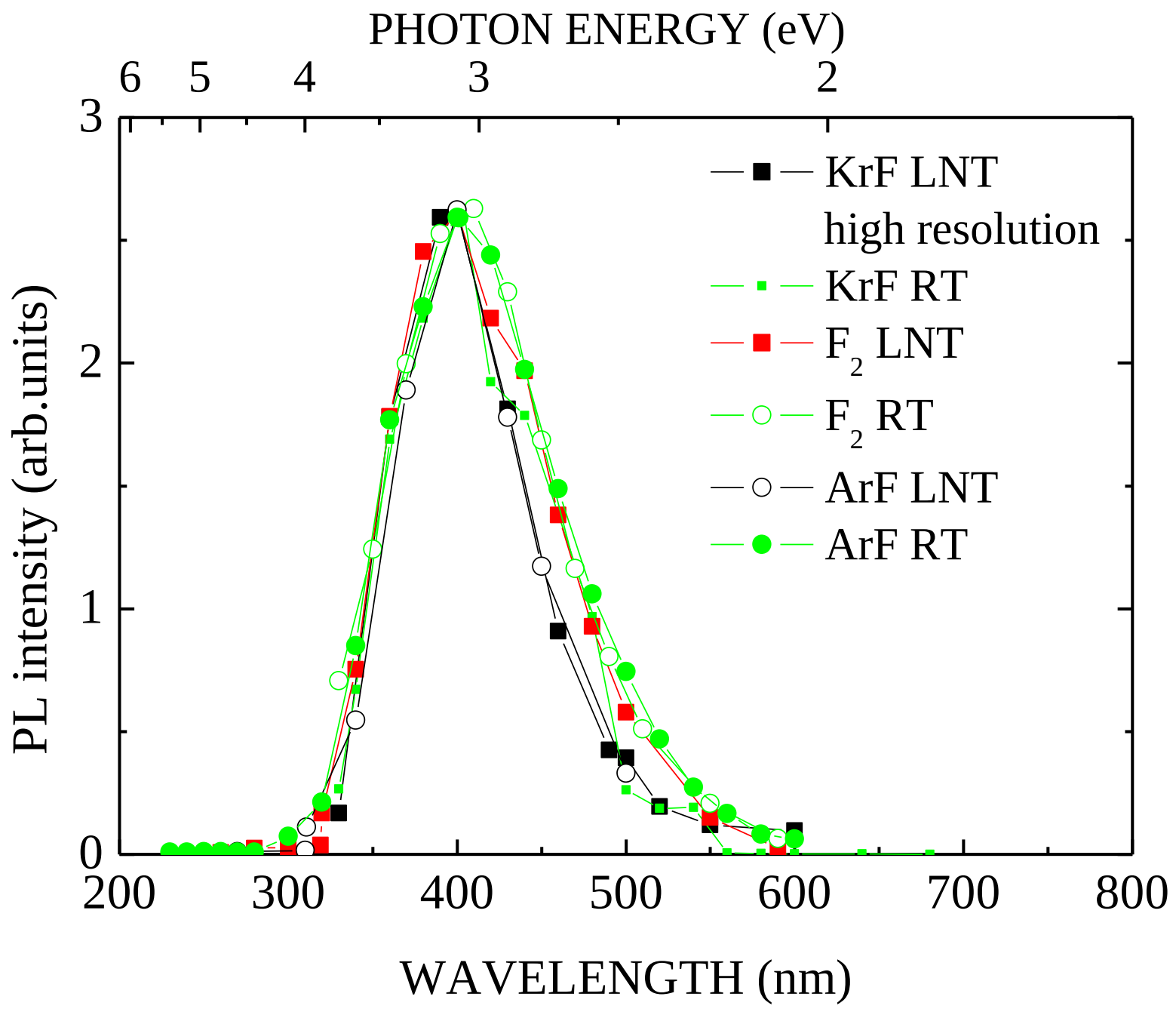

**Figure 1.** Time-resolved photoluminescence spectra of stishovite single crystal under excimer laser excitation at 293 K and 80 K. Each point corresponds to the integrated decay kinetics curve. The curves have been measured with the photon counting method.

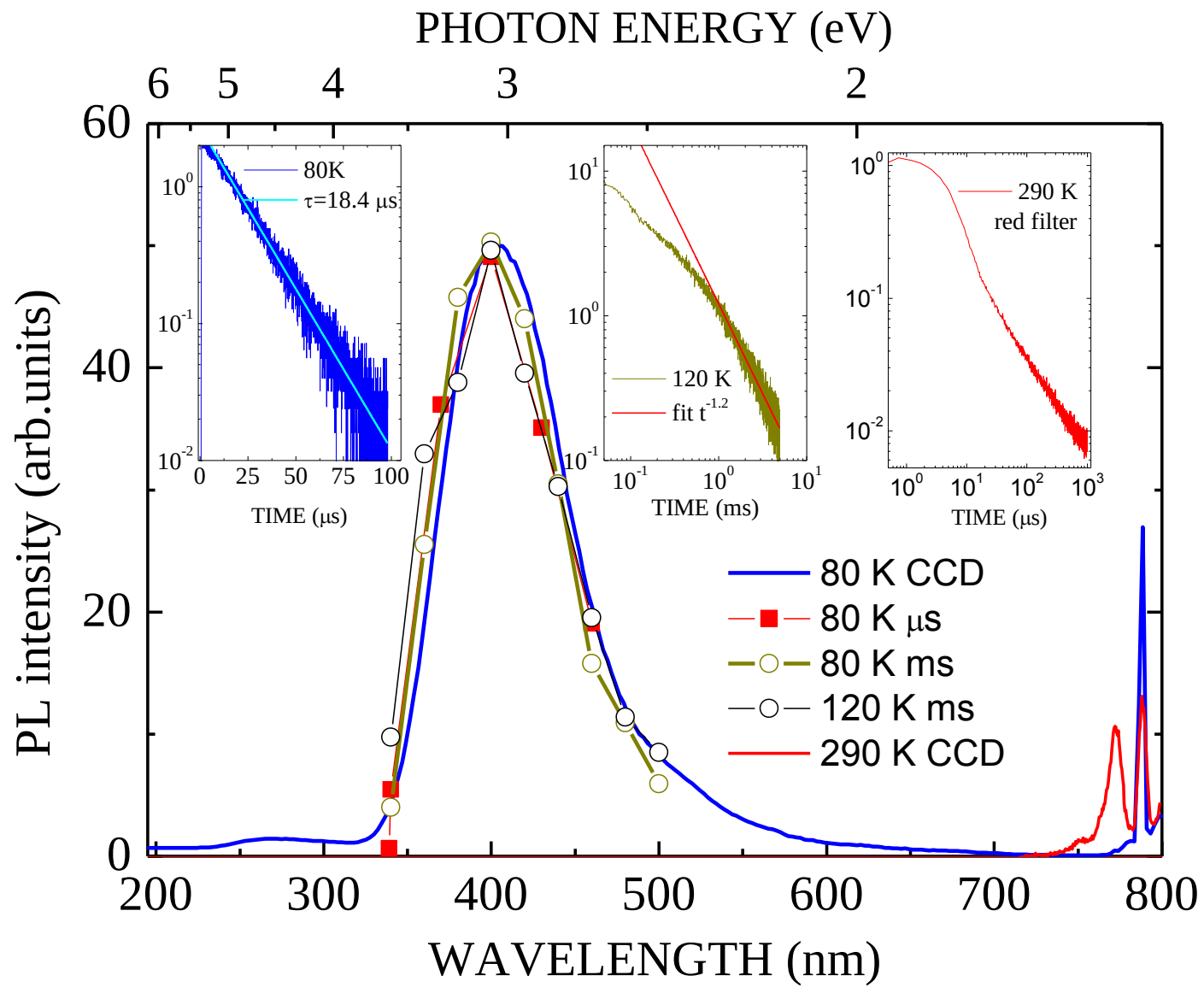

**Figure 2.** Time-resolved photoluminescence spectra of stishovite single crystal under pulses of excimer ArF laser excitation at 120 K and 80 K. Decay curves were obtained with PM in the current regime for µs time range and with photon counting module for ms time range. Steady state spectra have been measured with CCD of Hamamatsu minispectrometer. Inserts – the decay kinetic curves. Left – for the µs time range, middle and right – for the ms time range. The right insert shows decay kinetics measured in the range of sharp luminescence lines ascribed to carbon related molecular centers.

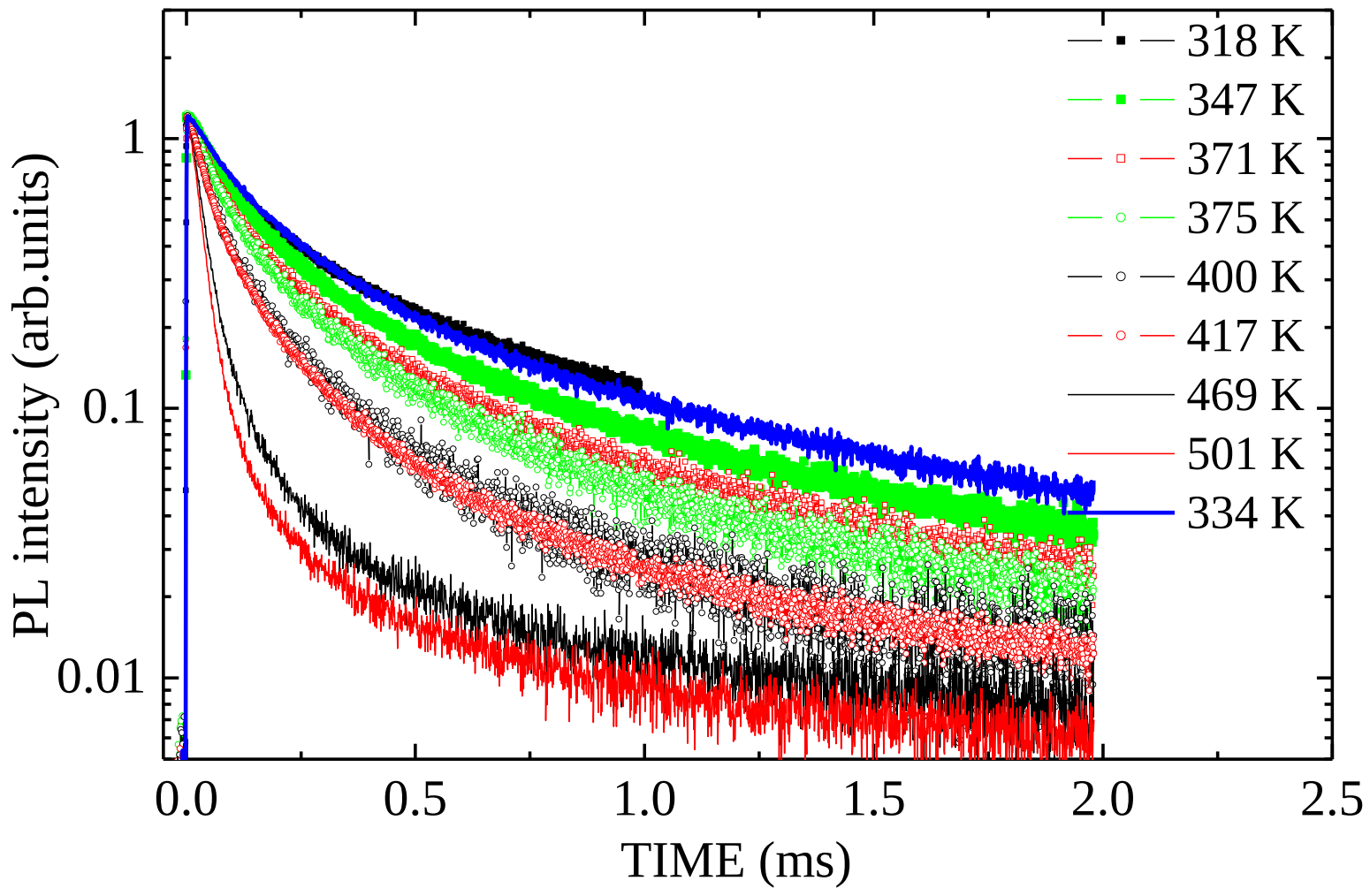

**Figure 3.** Photoluminescence decay curves of stishovite single crystal excited with pulses of ArF excimer laser in the range of temperatures above 300 K measured with the photon counting method.

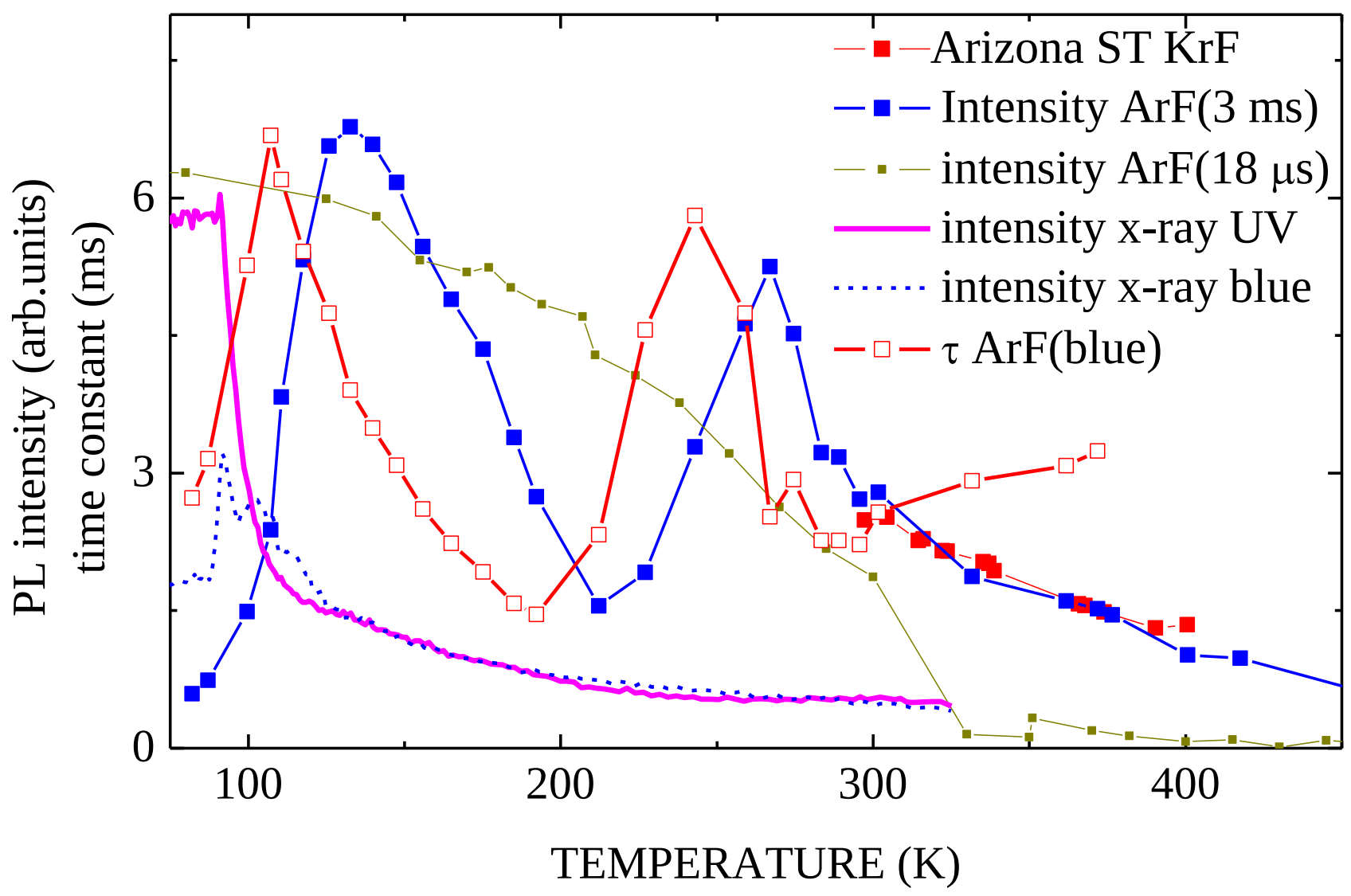

**Figure 4.** Time-resolved temperature dependences of the blue photoluminescence of stishovite measured with the photon counting method in the time range of 3 ms (big closed squares for intensity and open squares for time constant) and in the time range of 18 μs (small closed squares) under ArF laser excitation. Some points with the closed squares were measured for Arizona "natural" stishovite with the use of KrF laser. The lines correspond to x-ray excited luminescence intensity thermal dependences (line – UV band, dash line – blue band).

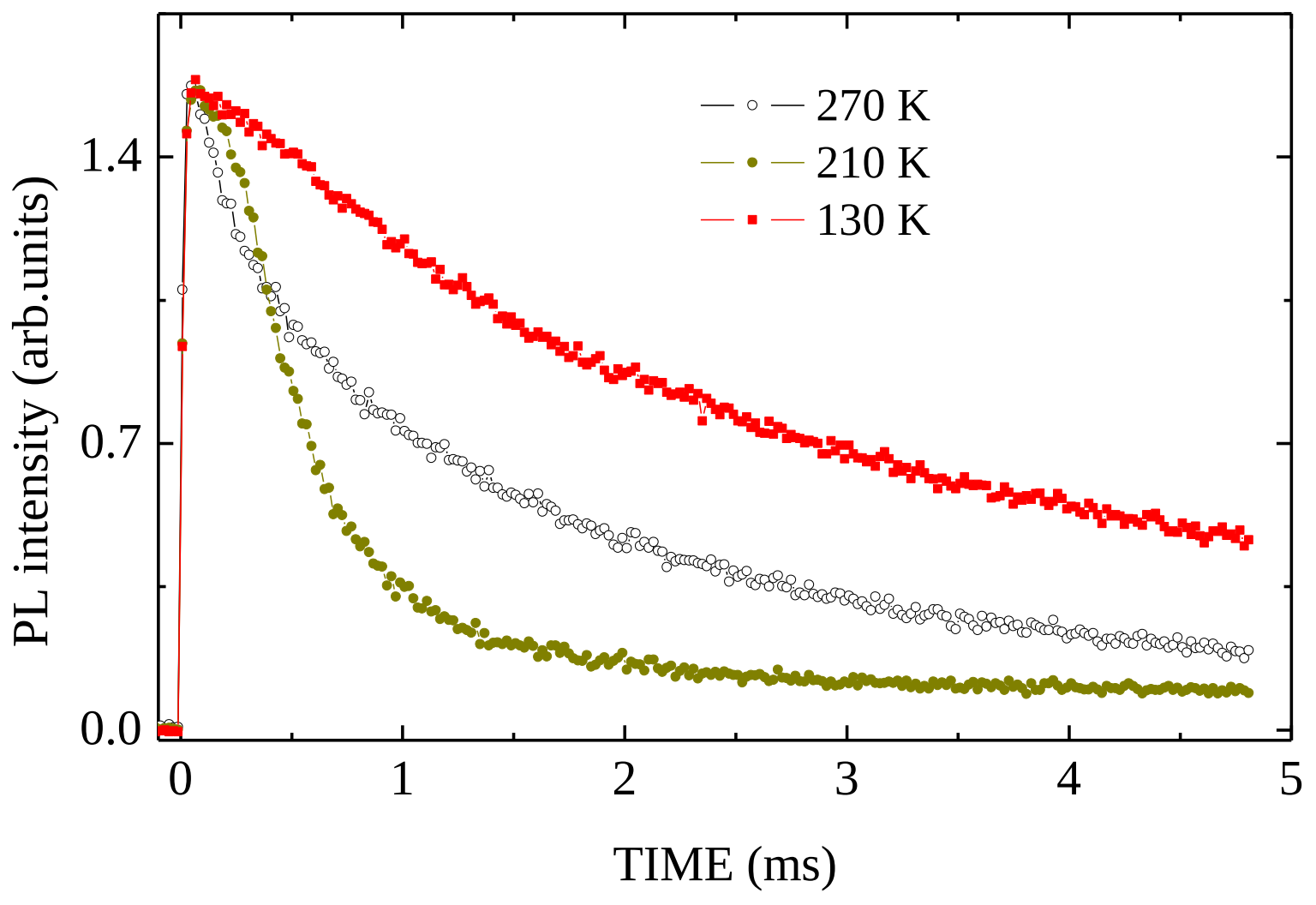

**Figure 5.** Slow blue PL decay curves measured at temperatures of maxima (130 K and 270 K) and minimum (210 K) of the dependence presented in Fig.4 for stishovite single crystal excited with ArF laser.

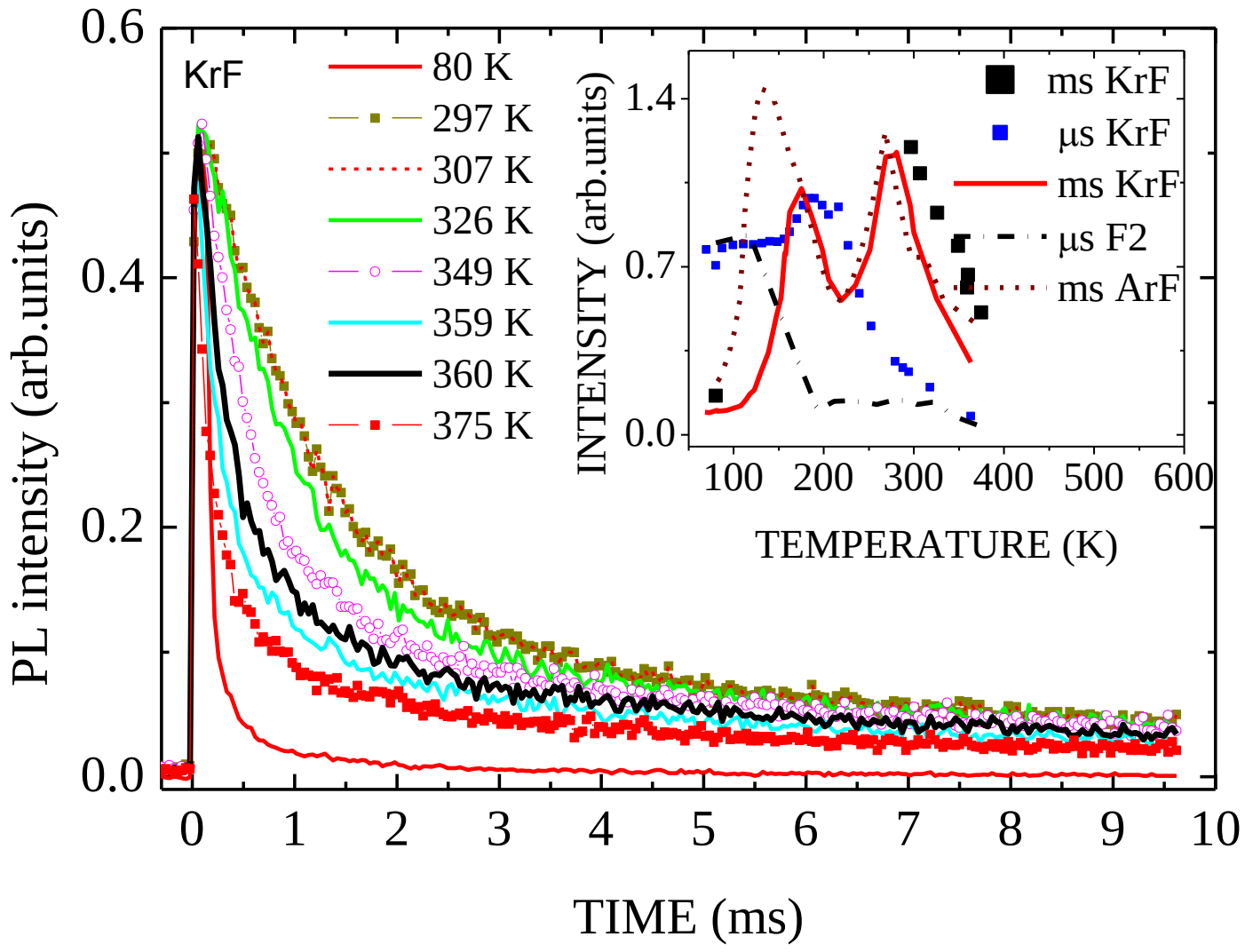

**Figure 6.** Photoluminescence decay curves of stishovite single crystal excited with pulses of KrF excimer laser in the 80 - 375 K range of temperatures measured with the photon counting method. Insert – comparison of thermal dependences obtained with the use of different excimer lasers in the ms and µs range of time.

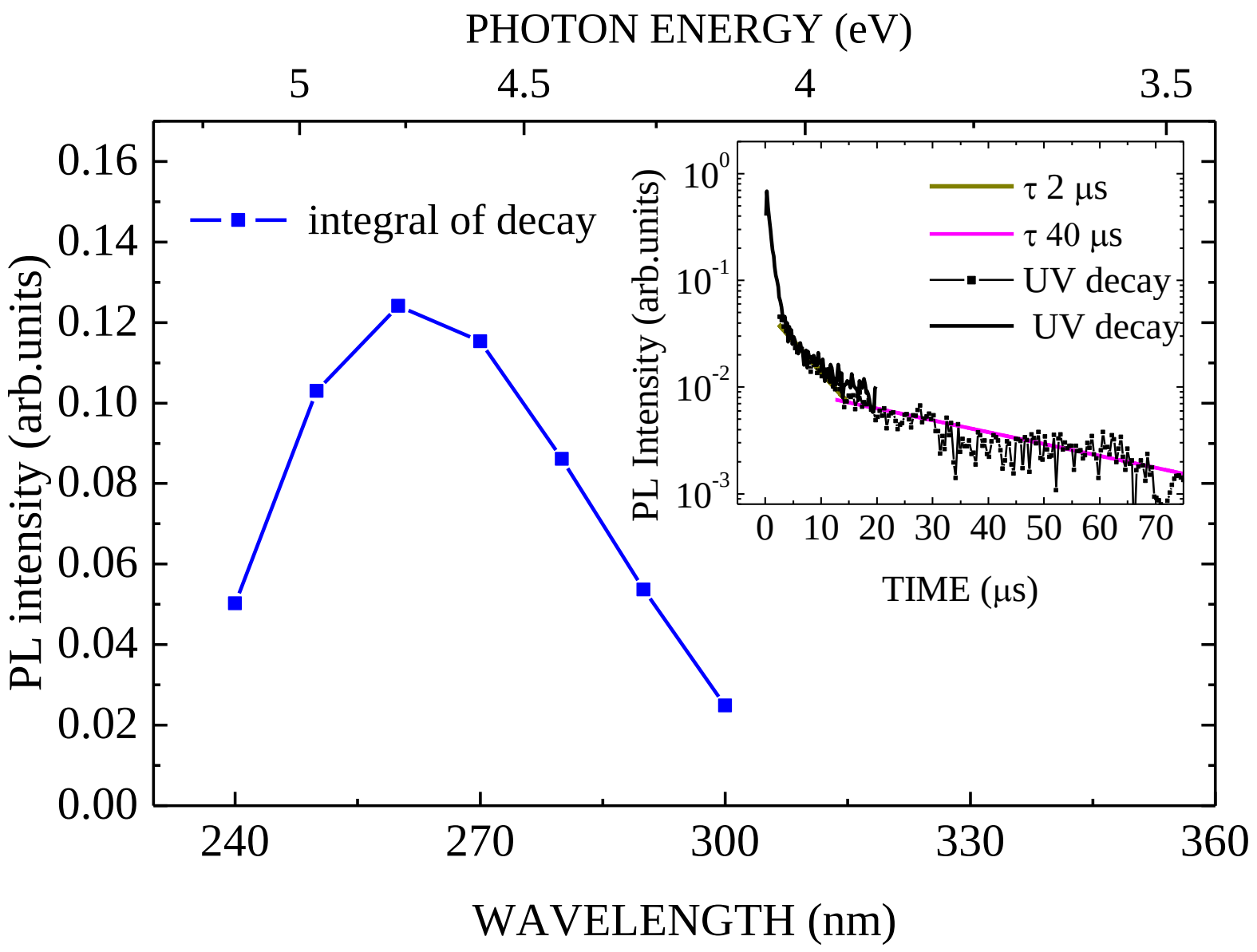

**Figure 7.** Time-resolved luminescence spectra measured with the photon counting method for the UV band of stishovite single crystal. Excitation – ArF laser. Insert: measured decay curve and exponential approximation with two exponents at T = 290 K.

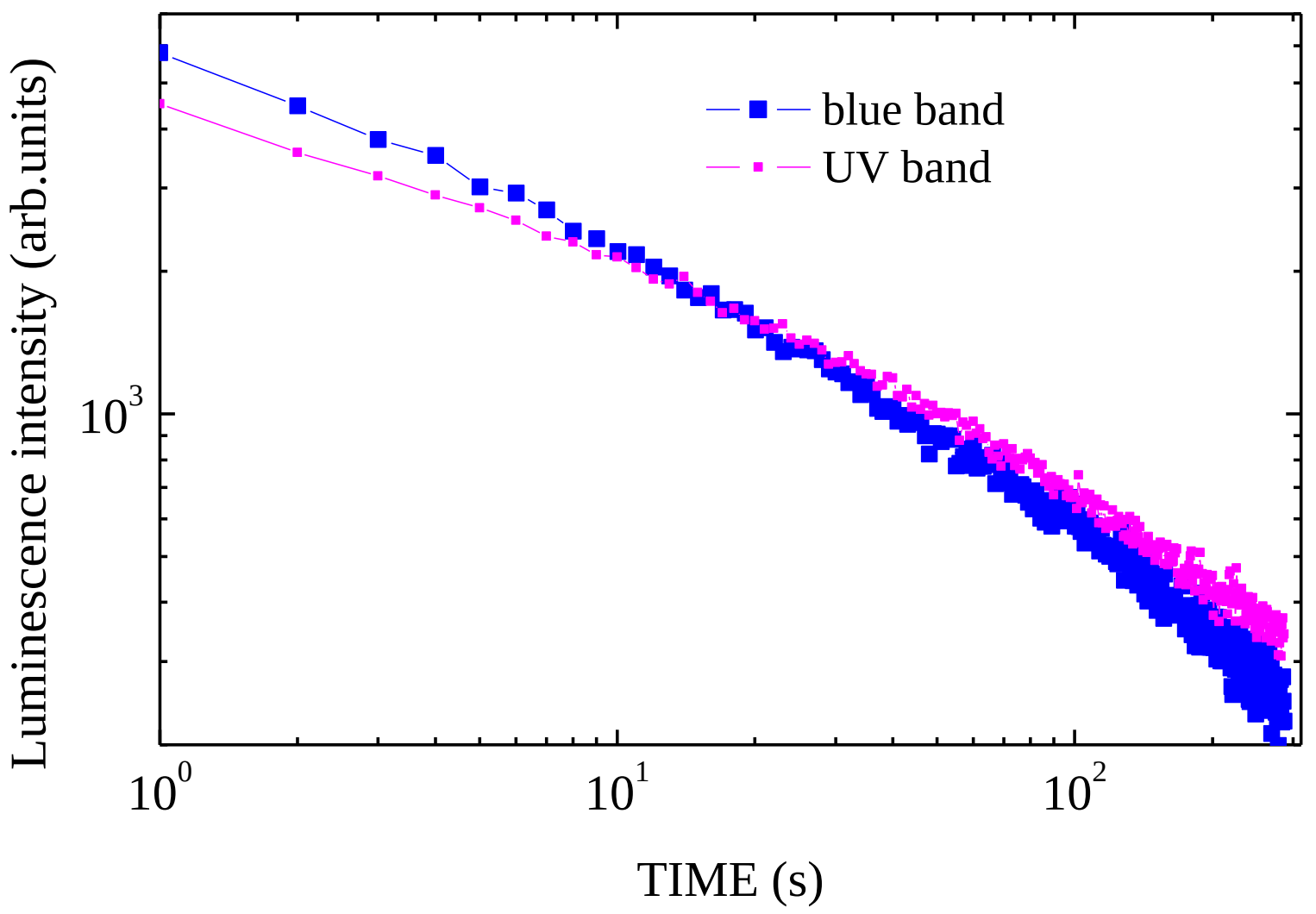

**Figure 8.** Afterglow kinetics after x-ray irradiation. Big closed squares – UV band and small closed squares - blue band. T = 80 K.

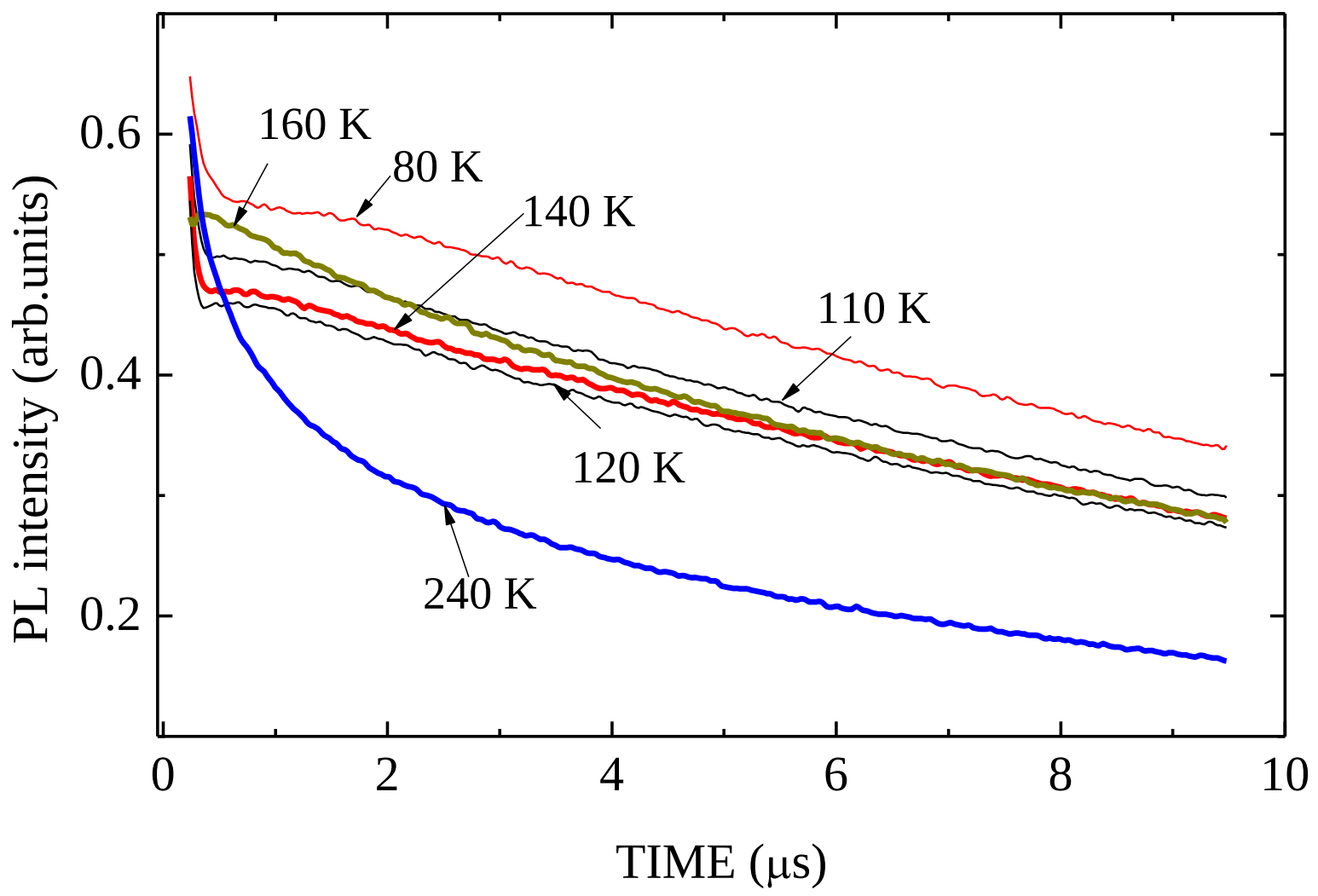

**Figure 9.** Blue PL decay kinetic curves of stishovite single crystal excited with ArF excimer laser at different temperatures. PM in current regime.

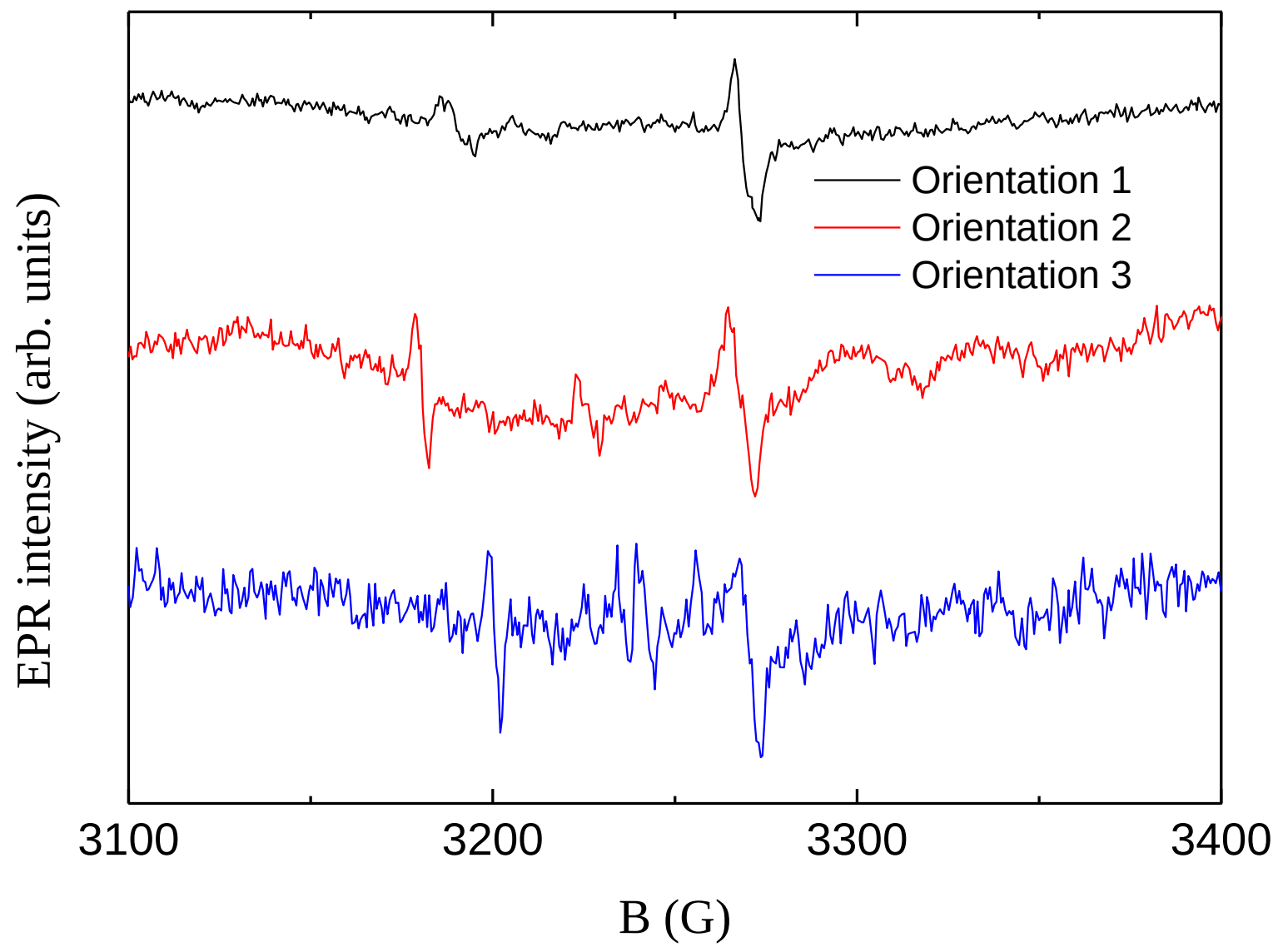

**Figure10.** X-band ESR spectra of stishovite at selected crystal orientations.

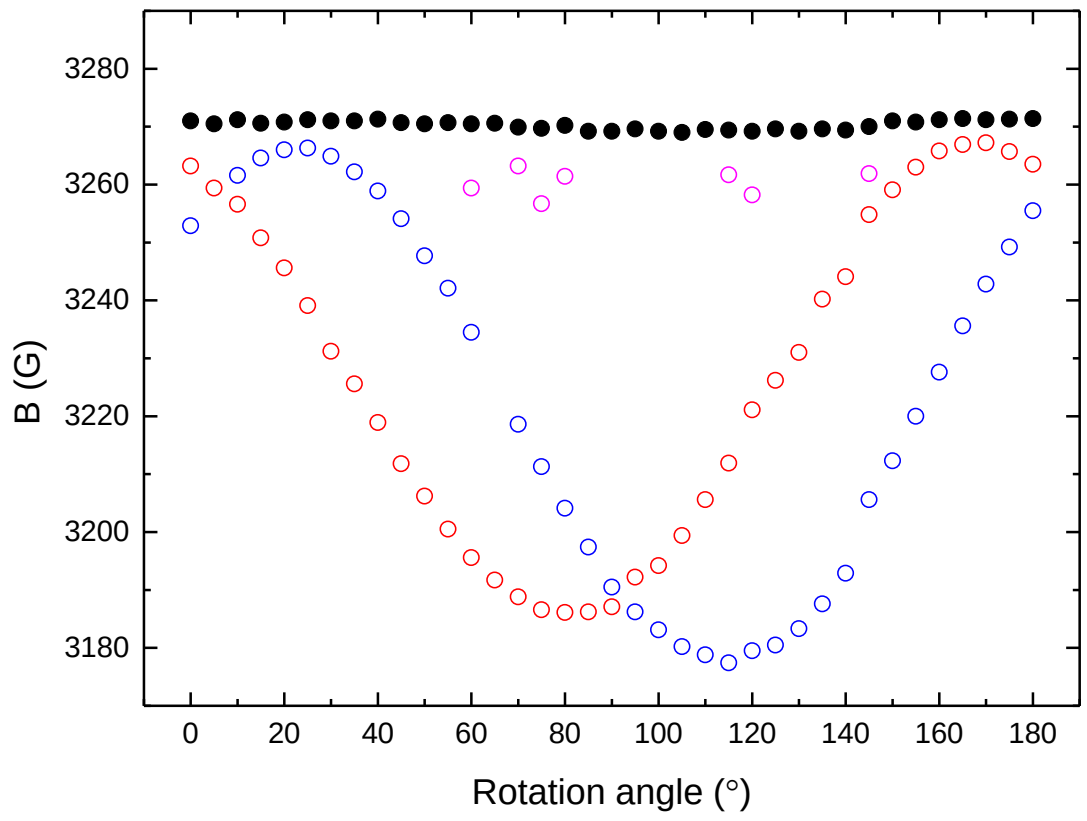

**Figure11.** The ESR resonance position roadmap of stishovite at 77 K around an arbitrary rotation axis. The filled circles originate from the sample holder.